\documentclass[sigconf,authorversion]{acmart}

\usepackage[normalem]{ulem}
\usepackage{multirow}

\AtBeginDocument{%
  \providecommand\BibTeX{{%
    \normalfont B\kern-0.5em{\scshape i\kern-0.25em b}\kern-0.8em\TeX}}}

\copyrightyear{2023} 
\acmYear{2023} 
\setcopyright{acmlicensed}\acmConference[CIKM '23]{Proceedings of the 32nd ACM International Conference on Information and Knowledge Management}{October 21--25, 2023}{Birmingham, United Kingdom}
\acmBooktitle{Proceedings of the 32nd ACM International Conference on Information and Knowledge Management (CIKM '23), October 21--25, 2023, Birmingham, United Kingdom}
\acmPrice{15.00}
\acmDOI{10.1145/3583780.3615132}
\acmISBN{979-8-4007-0124-5/23/10}

\usepackage{booktabs} 
\usepackage{color, soul}
\usepackage{enumitem}
\usepackage{balance}

\definecolor{lightgreen}{RGB}{190,230,219}
\DeclareRobustCommand{\hlgreen}[1]{{\sethlcolor{lightgreen}\hl{#1}}}

\newcommand{\captionshrink}{\vspace*{-0.5\baselineskip}}

\begin{document}

\fancyhead{}
\title{Towards Filling the Gap in Conversational Search: From Passage Retrieval to Conversational Response Generation}

\author{Weronika Łajewska}
\affiliation{%
  \institution{University of Stavanger}
  \city{Stavanger}
  \country{Norway}
}
\email{weronika.lajewska@uis.no}

\author{Krisztian Balog}
\affiliation{%
  \institution{University of Stavanger}
  \city{Stavanger}
  \country{Norway}
}
\email{krisztian.balog@uis.no}

\begin{abstract}
Research on conversational search has so far mostly focused on query rewriting and multi-stage passage retrieval. However, synthesizing the top retrieved passages into a complete, relevant, and concise response is still an open challenge. Having snippet-level annotations of relevant passages would enable both (1) the training of response generation models that are able to ground answers in actual statements and (2) the automatic evaluation of the generated responses in terms of completeness. 
In this paper, we address the problem of collecting high-quality snippet-level answer annotations for two of the TREC Conversational Assistance track datasets.
To ensure quality, we first perform a preliminary annotation study, employing different task designs, crowdsourcing platforms, and workers with different qualifications. Based on the outcomes of this study, we refine our annotation protocol before proceeding with the full-scale data collection. Overall, we gather annotations for 1.8k question-paragraph pairs, each annotated by three independent crowd workers.
The process of collecting data at this magnitude also led to multiple insights about the problem that can inform the design of future response-generation methods.
This is an extended version of the article published with the same title in the Proceedings of CIKM'23.

\end{abstract}



\maketitle

\section{Introduction}
\label{sec:intro}

A large fraction of research on conversational information seeking (CIS) to date has focused on the problem of retrieving relevant passages. The Conversational Assistance track at the Text Retrieval Conference (TREC CAsT)~\citep{Dalton:2019:TREC, Dalton:2020:TREC, Dalton:2021:TREC, Owoicho:2022:TREC} has played a major role in enabling research on this task by developing a series of reusable test collections. The task of conversational passage retrieval requires advances in query rewriting~\citep{Lin:2021:TOIS,Vakulenko:2021:ECIR,Vakulenko:2021:WSDM} and can also directly benefit from research on multi-stage passage retrieval~\citep{Luan:2021:MIT}. However, identifying relevant passages is only an intermediate step. Ultimately, the information contained in these passages would need to be synthesized into a single answer. \emph{Conversational response generation} is the task of encapsulating the most relevant pieces of information in an easily consumable unit~\citep{Culpepper:2018:SIGIRForum}. Including it in the CIS pipeline would increase the naturalness of the conversation~\citep{Trippas:2019:IPM, Trippas:2018:CHIIR}.

There are at least two main challenges involved in the task of response generation: identifying key pieces of information from relevant results (e.g., paragraphs) and summarizing them in a concise answer. Correspondingly, \citet{Ren:2021:ToIS} propose to split the task into two stages: (1) identification of supporting snippets and (2) summarization of selected snippets. In this paper, we focus on the problem of (1), and more specifically on building a snippet dataset with high-quality annotations using crowdsourcing.

The significance of being able to identify relevant snippets is twofold. 
First, it enables the training of models that can ground the generated answers in actual statements. Natural language generation models are susceptible to hallucinations, especially if the query is insufficiently covered in the corpus, or the retrieved documents contain redundant, complementary, or contradictory information~\citep{Ji:2022:arxiv}. Therefore, employing abstractive summarization methods on top of relevant snippets identified can help to mitigate this problem and provide more control over the generation process, much in the spirit of the two-step process proposed in~\citep{Ren:2021:ToIS}.
Second, it would enable automatic evaluation of the generated responses quantitatively, in terms of relevant information nuggets included~\citep{Pavlu:2012:WSDM}.
Response summarization in CIS systems has been piloted in the most recent edition of TREC CAsT~\citep{Owoicho:2022:TREC}, where the quality of answer summaries is evaluated by human judges along three dimensions: relevance, naturalness, and conciseness~\citep{Owoicho:2022:TREC}. 
Having annotations of relevant snippets would enable automatic evaluation of answers in terms of completeness.

Even though crowdsourcing has become an established means of collecting human annotations at scale, ensuring data quality can be challenging~\citep{Daniel:2018:CSUR}.
Indeed, we demonstrate that the seemingly straightforward task of highlighting relevant snippets may not be so simple and deserves more close attention.

In this paper, we first investigate what are effective task designs and trade-offs between worker qualifications and costs to perform the task of snippet annotations. Specifically, we consider paragraph- and sentence-level snippet annotation interfaces, multiple crowdsourcing platforms, and crowd workers with different qualifications as well as expert annotators.
Measuring the quality of annotations is challenging because relevant snippet selection is subjective and often there are multiple correct sets of snippets in a given passage. We evaluate the resulting annotations in terms of inter-annotator agreement and similarity to expert annotations using text similarity measures adapted to this task.

Based on the results of our preliminary study, we set out to create a large-scale dataset, CAsT-snippets, which enriches the TREC CAsT 2020 and 2022 datasets with snippet-level answer annotations.
We follow a setup in which we closely work with a selected pool of highly engaged crowd workers in order to ensure high data quality.
Our findings from this data collection effort reveal numerous associated challenges that can help inform the design of response generation methods in future work.

The resources developed in this study (annotated data and code for computing evaluation measures) are made publicly available at \url{https://github.com/iai-group/CAsT-snippets}.
\section{Related work}
\label{sec:related_work}

Research on conversational response generation has attracted a lot of attention in task-oriented dialogue systems~\citep{Budzianowski:2018:EMNLP, Lippe:2020:arXiv, Pei:2019:ECAI}, question answering~\citep{Baheti:2020:ACL}, open-domain chatbots~\citep{Xing:2016:AAAI,Dziri:2019:ACL}, and most recently in conversational information seeking, as part of TREC CAsT'22~\citep{Owoicho:2022:TREC}. 
The performance of response generation is commonly evaluated using automatic similarity measures for natural language generation tasks such as BLEU~\citep{Papineni:2002:ACL, Kocisk:2017:TACL}, and ROUGE~\citep{Lin:2004:ACL}. However, some dimensions are not reliably covered by currently available automatic metrics and require manual evaluation (e.g., coherence and relevance)~\citep{Fabbri:2020:ACL}, while others (e.g., completeness) can be evaluated automatically, provided that more fine-grained annotations are available.
Information nuggets, defined as \emph{minimal, atomic units of relevant information} of retrieved documents, have been proposed as an alternative to automatically assign relevance judgments to documents and/or evaluate retrieval systems~\citep{Pavlu:2012:WSDM}.
Our work aims to contribute to this type of evaluation by studying ways to collect snippet-level annotations.
A task similar to snippet annotation (or information nuggets identification) has been broadly researched in QA systems. In most available datasets for reading comprehension focused mainly on factoid questions, the generated response is a single entity or a short segment of text from the passage~\citep{Rajpurkar:2016:EMNLP, Campos:2016:arXiv, Tan:2017:arXiv, Choi:2018:EMNLP}.

Crowdsourcing provides a scalable means to the completion of large amounts of labeling or annotation tasks that require human intelligence~\citep{Gadiraju:2015:IEEE}.
The actual quality of the results is influenced by the workers, software platform~\citep{Vakharia:2013:arXiv}, task design~\citep{Eickhoff:2018:WSDM}, and quality measures employed~\citep{Daniel:2018:CSUR}. 
This paper attempts to understand what setup is needed to effectively perform the task of snippet annotation. 

Relevant annotations efforts include QuaC~\citep{Choi:2018:EMNLP}, which is a dataset of QA dialogues. 
However, it is limited to sections of Wikipedia articles and contains only dialogues about a biased sample of entities of type person. Queries in CAsT datasets are much more diverse, both in terms of the expected type of answer and in the topics discussed.
Most relevant to our paper is the work by \citet{Ren:2021:ToIS}, where crowd workers are asked to respond to queries from the TREC CAsT'19 dataset while being presented with SERPs.
The response generation task is divided into three stages: (optional) query rewriting, finding supporting sentences in results displayed on a SERP, and summarizing them into a short conversational response.
We focus only on the supporting evidence finding step, which is performed on a finer (snippet-level) granularity, and explore various task designs to ensure high data quality. 

\section{Dataset}
\label{sec:dataset}

We perform annotations on the TREC CAsT 2020 and 2022 datasets.\footnote{ The 2019 dataset has relatively low complexity compared to these two, while the 2021 dataset provides relevance assessments on the level of documents instead of passages.}
Each dataset comprises of a set of information-seeking dialogues (i.e., topics) with a sequence of questions (i.e., queries) within each.
The input to the snippet annotation task consists of queries and corresponding passages. 
We consider the top 5 passages for each query with respect to their relevance labels in the ground truth (ranging from 0 to 4). If there are fewer than 5 passages available for the query at the highest relevance level, then we fill up the remaining slots with passages one relevance level below. If there are more passages available, then we cluster them using \emph{k}-means clustering and pick a random passage per cluster.  For example, if we have 3 highly relevant passages for a given query and 10 relevant passages, we choose all the passages with relevance level 4 and populate the remaining two places by splitting the passages with a relevance level 3 into two clusters and then choosing a random passage from each cluster. 
Selecting the passages for annotation this way ensures that they are both relevant and diverse. Even though we mostly consider highly relevant and relevant passages, some of them do not contain a direct answer to the question, which makes the snippet annotation task even more challenging.
\section{Evaluation measures}
\label{sec:evaluation_measures}

\begin{figure}
    \centering
\includegraphics[width=0.4\textwidth]{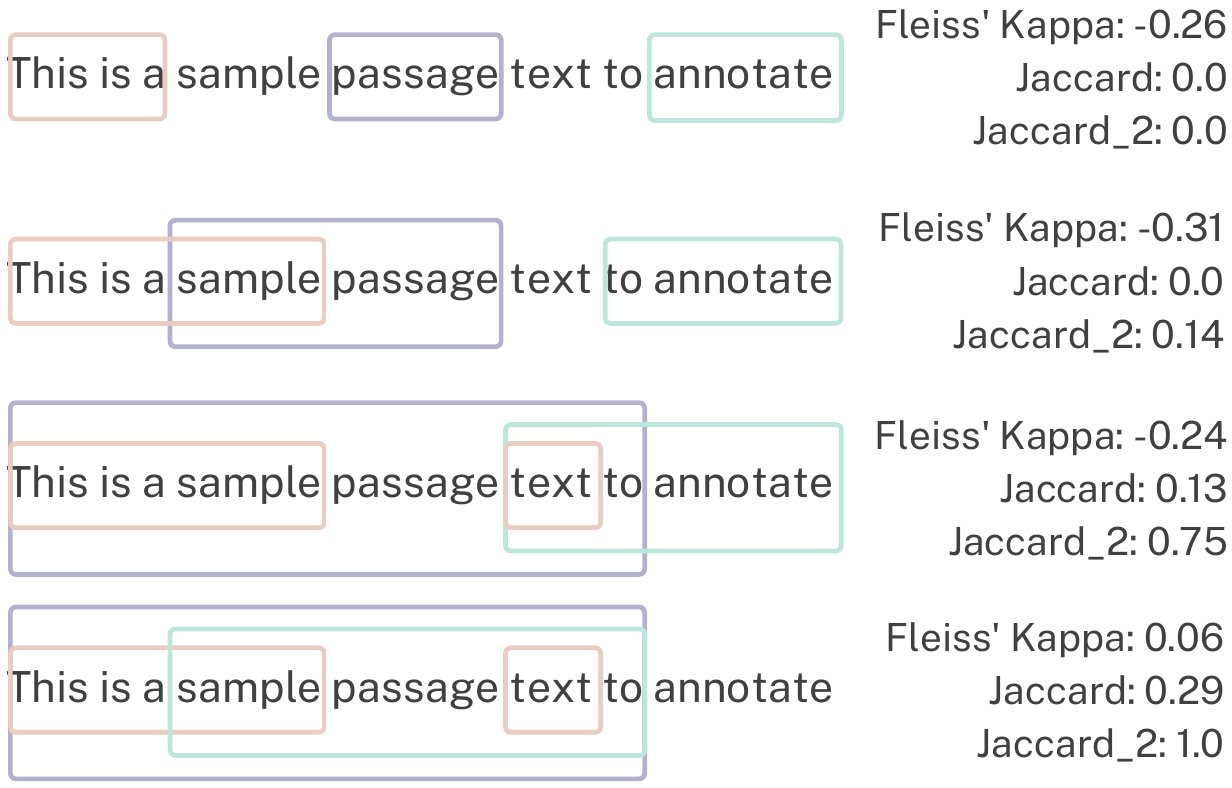}
    \caption{Visualization of annotations made by 3 workers on a sample text. The values of Fleiss' Kappa indicate poor agreement. On the other hand, Jaccard similarity offers more granular results which are easier to interpret in this specific scenario. Krippendorf's Alpha with nominal weight function gives analogous results to Fleiss' Kappa.} 
    \label{fig:annotations_visualization}
\end{figure}

Traditional metrics for inter-annotator agreement such as Fleiss' Kappa or Krippendorff's Alpha are designed to assess categorical annotations and rely on a binary notion of agreement. In our case, we are more interested in measuring the degree to which snippets selected by different workers overlap (see Fig.~\ref{fig:annotations_visualization}). We define evaluation measures to compare the agreement between annotators and across crowd workers and expert annotators.

\subsection{Inter-annotator Agreement}

We define inter-annotator agreement in terms of Jaccard similarity.  Given an input text $t$ annotated by $n$ workers ($w_1, \dots, w_n$), we count the length of the snippets chosen by all annotators and divide it by the length of snippets chosen by any annotator.  
Formally:
\begin{equation}
    J(t) = \frac{|\bigcap_{i=1}^n snippets(t,w_i)|}{|\bigcup_{i=1}^n snippets(t,w_i)|} ~, \label{eq:jaccard}
\end{equation}
where $snippets(t,w_i)$ denotes the set of intervals selected by worker $w_i$ in text $t$. The intersection and union of snippet intervals is calculated on the character level.

We also consider a less strict variant of the measure, termed Jaccard$_k$, which takes only those intervals into account that are chosen by at least $k$ annotators.  Formally, in the nominator in Eq.~\eqref{eq:jaccard} we count the length of intervals that appear in at least $k$ annotations made by different workers, while the denominator remains unchanged. 

\subsection{Similarity to Reference Annotations}

To measure the similarity of snippet annotations by crowd workers against reference annotations by experts, we follow a logic similar to ROUGE-1, which considers the overlap of unigrams between the system and reference summaries~\citep{Lin:2004:ACL}. Specifically, we employ the ROUGE-like measures proposed in~\citep{Iskender:2021:NLDB}.  For every input text $t$, we have annotations made by $n$ different crowd workers ($w_i$) and reference annotations by $m$ different experts ($e_j$).
First, we define precision and recall of the snippets in text $t$ between a pair of annotators $w_i$ and $e_j$:
\begin{eqnarray*}
    p_t^{i,j}= \frac{|snippets(t,w_i) \cap snippets(t,e_j)|}{|snippets(t,w_i)|} ,  \\
    r_t^{i,j}= \frac{|snippets(t,w_i) \cap snippets(t,e_j)|}{|snippets(t,e_j)|} . 
\end{eqnarray*}
We compute the F1 score as the harmonic mean of precision and recall: $f1_t^{i,j}= 2\times p_t^{i,j} \times r_t^{i,j}/(p_t^{i,j} + r_t^{i,j})$.

Next, we aggregate these measures for a given crowd worker $i$ against all ($m$) expert annotations:
precision as $P_t^{i} = \frac{1}{m}\sum_{j=0}^{m}{p_t^{i,j}}$,
recall as $R_t^{i} = \frac{1}{m}\sum_{j=0}^{m}{r_t^{i,j}}$,
and F1 score as $F1_t^{i} = \frac{1}{m}\sum_{j=0}^{m}{f1_t^{i,j}}$.

Finally, we aggregate the annotations across all ($n$) crowd workers in three different ways:
\begin{itemize}[leftmargin=0.5cm]
    \item \emph{Mean} ($\overline{P}_t, \overline{R}_t, \overline{F1}_t$), by simply averaging $P_t^{i}$, $R_t^{i}$, and $F1_t^{i}$ over all crowd workers.
    \item \emph{Majority} ($P_t^\gg, R_t^\gg, F1_t^\gg$), where we consider a single crowd worker snippet annotation, which is taken as the union of intervals that are selected by the majority of workers.
    \item \emph{Similarity} ($P_t^\simeq, R_t^\simeq, F1_t^\simeq$), where we only consider the snippet annotation by crowd worker $w_i$ that is most similar to the annotations of other crowd workers in terms of $f1_t^{i,j}$.
\end{itemize}
\section{Preliminary study}
\label{sec:preliminary_study}

\begin{figure}
    \centering
    \scriptsize
    \begin{tabular}{p{8cm}}
        \textbf{Query:} I remember Glasgow hosting COP26 last year, but unfortunately I was out of the loop. What was the conference about? \\
        \vspace{0.1cm}
        \textbf{Passage:} HOME - UN Climate Change Conference (COP26) at the SEC – Glasgow 2021 Uniting the world to tackle climate change. The UK will host the 26th UN Climate Change Conference of the Parties (COP26) in Glasgow on 1 – 12 November 2021. The COP26 summit will \hlgreen{bring parties together to accelerate action towards the goals of the Paris Agreement and the UN Framework Convention on Climate Change}. The UK is committed to working with all countries and joining forces with civil society, companies and people on the frontline of climate change to inspire climate action ahead of COP26. COP26 @COP26 · May 25, 2021 1397069926800654339 We need to accelerate the \#RaceToZero Join \@wef, \@MPPindustry, \@topnigel \& \@gmunozabogabir for \hlgreen{a series of events demonstrating the need for systemic change to accelerate the global transition to net zero}. Starting May 27th Learn more \#ClimateBreakthroughs | \#COP26 Twitter 1397069926800654339 COP26 \@COP26 · May 24, 2021 1396737733649846273 \hlgreen{\#TechForOurPlanet is a new challenge programme for \#CleanTech startups to pilot and showcase their solutions at \#COP26}! Innovators can apply to six challenges focusing around core climate issues and government priorities.
    \end{tabular}
    \caption{Sample from the CAsT-snippets dataset with highlighted expert annotations.}
    \label{fig:dataset_sample}
\end{figure}

To ensure that we get high-quality snippet-level annotations, we first perform a preliminary study where we compare different task designs, platforms, and worker pools, by annotating two topics selected from the TREC CAsT'22 dataset, with markedly different characteristics, comprising 22 queries in total.
The first topic (ID 132) has 12 turns (i.e., queries) and focuses on listing several independent pieces of information, which requires workers to choose multiple keywords or phrases within each passage. The second topic (ID 133) consists of 10 turns with questions about recipes, where several consecutive sentences contain relevant bits of information to be included in the answer.

\subsection{Task Designs}

\begin{figure*}[!t]
    \centering
    \includegraphics[width=0.8\textwidth]{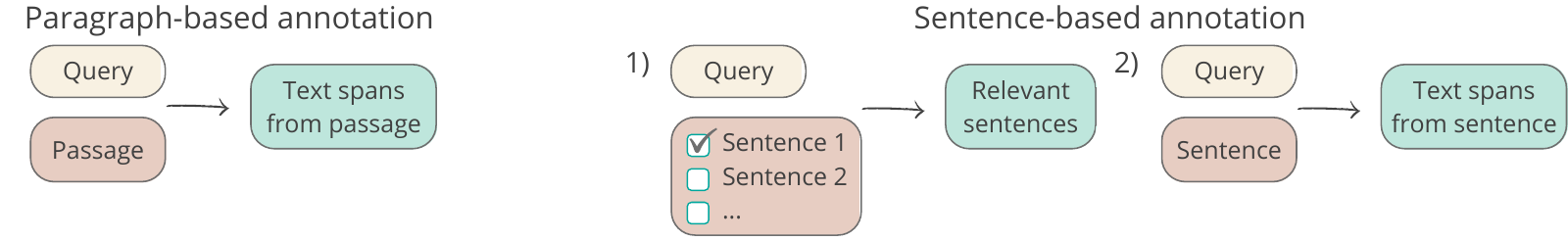}
    \captionshrink
    \caption{Illustration of different designs for the snippet annotation task.}
    \label{fig:tasks_schemas}
\end{figure*}

We task crowd workers with the identification of snippets in a provided text that contains key pieces of the answer to a given query. Text snippets are required to be short, concise, informative, self-contained, and cannot overlap. Each snippet is supposed to contain one piece of information, so it can be treated as an information nugget. An example expert-annotated snippet is shown in Fig.~\ref{fig:dataset_sample}. Specifically, we identify snippets in paragraphs that have been labeled as relevant answers to the question. These passages can be long, which makes the annotation task cognitively demanding. Therefore, we consider two designs of the task:  paragraph-based and sentence-based; see~Figure~\ref{fig:tasks_schemas}. 

In the \emph{paragraph-based} annotation task, workers are asked to identify all text snippets in a given passage that are relevant to the input query. Since paragraphs can be lengthy, we also consider a simplified, \emph{sentence-based} variant of this task, which lets workers operate on the significantly shorter text and enforces shorter text snippet selection. Specifically, the task is divided into: (1) relevant sentence selection, and (2) snippet annotation in relevant sentences. In sub-task (1), crowd workers are presented with a question and a passage that is split into sentences. They are asked to choose sentences that contain information relevant to the query. This is a straightforward task that resembles extractive summarization~\citep{Zhou:2018:ACL}. Sub-task (2) is very similar to the paragraph-based annotation task, the only difference is that workers are presented with a relevant sentence instead of an entire passage. 

\subsection{Platforms and Workers}

We set up the annotation task on two crowdsourcing platforms: Amazon MTurk\footnote{\url{https://www.mturk.com/}} and Prolific.\footnote{\url{https://app.prolific.co/}} MTurk offers an easily customizable web-based annotation interface and it is possible to filter workers based on qualifications. Prolific has more limited options in terms of the annotation interface, but the qualification of workers is {claimed to be higher than on MTurk.\footnote{\url{https://www.prolific.co/prolific-vs-mturk}} Additionally, we employ a group of expert annotators (Ph.D. students) who have been trained to perform this annotation task; they also use the MTurk platform, but in sandbox mode, i.e., without receiving payment. 

\subsection{Task Configurations}

\begin{table}
    \centering
    \small
    \caption{Task configurations used for data collection; values are averaged for the annotation of a single paragraph.}
    \captionshrink
    \hspace*{-0.1cm}
    \begin{tabular}{l|l|r|c|c|c}
        \begin{tabular}  {@{}l@{}}\textbf{Task}\\\textbf{Variant}\end{tabular} & \textbf{Annotator} & \textbf{Time} & \begin{tabular}   {@{}l@{}}\textbf{\#}\\\textbf{workers}\end{tabular} & \begin{tabular}   {@{}l@{}}\textbf{Acceptance}\\\textbf{rate}\end{tabular} & \textbf{Cost} \\
        \hline
        \multirow4{*}{Paragraph} & MTurk regular & 182s & 5 & 50\% & \$0.36 \\
        & MTurk master & 63s & 5 & 90\% & \$0.38 \\
        \cline{2-5}
        & Prolific & 154s & 5 & 79\% & \$0.51 \\ 
        \cline{2-5}
        & Expert & 96s & 3 & - & - \\ 
        \hline
        \multirow{2}{*}{Sentence} & MTurk regular & 977s & 3 & 72\% & \$0.43 \\
        & MTurk master & 305s & 3 & 87\% & \$0.56 \\
    \end{tabular}
    \label{tab:task_stats}
\end{table}

Table~\ref{tab:task_stats} shows the different task configurations we experiment with. The paragraph-based annotation task, which is regarded as the cognitively more demanding variant, is performed with workers from both crowdsourcing platforms as well as with expert annotators. The sentence-based variant of the task is executed only on MTurk. All tasks on MTurk are performed with both regular and master workers.\footnote{MTurk Master is a qualification earned through a proven track record of quality work.} 
The remuneration varies depending on the platform used, with payments on MTurk ensuring that the payment is above the US federal minimum wage and on Prolific being determined by the recommended payment per minute suggested on the platform. The reported numbers correspond to the average cost of annotating one paragraph (based on the average number of sentences per paragraph in the case of the sentence-based variant) and also include the platform fee. The total cost of this preliminary study was \$1.2k. 
Table~\ref{tab:task_stats} also reports the number of workers assigned for each sample in the given task variant, the median time taken to annotate a paragraph, and the acceptance rate after a manual quality check. The acceptance rate is given only to tentatively present the difficulty of different variants of the task and it is not indicative of the quality of the final accepted annotations. 

\subsection{Quality Control}

In order to ensure that the collected data is of the highest quality, we define several automatic quality control criteria before the final manual verification of results. In the paragraph-based task variant, annotations longer than 50\% of the passage and annotations not contained in the intersection of the intervals chosen by at least two other crowd workers are flagged. In the first sub-task of the sentence-based variant, we flag submissions where fewer than one sentence or more than 75\% of sentences are chosen. Additionally, only submissions that have at least one common sentence chosen with other crowd workers are accepted. The second sub-task applies the same quality control criteria as the paragraph-based variant, with the maximum length of the snippet increased to 75\% of the sentence. Importantly, the automatic quality control criteria are only used to flag submissions that require additional attention. All results are manually verified by the first author of the paper, and responses that do not meet the task requirements are rejected.

\begin{table}
    \centering
    \small
    \caption{Inter-annotator agreements (J and J$\boldsymbol{_k}$). The number of annotators for every input text is shown in parentheses.}
    \captionshrink
    \hspace*{-0.1cm}
    \begin{tabular}{l|l|c|c|c|c}
        \multirow{2}{*}{\begin{tabular}   {@{}l@{}}\textbf{Task}\\\textbf{Variant}
        \end{tabular}} & \multirow{2}{*}{\textbf{Annotator}} & \multirow{2}{*}{\textbf{J}} & \multicolumn{3}{c}{\textbf{J}$\boldsymbol{_k}$} \\
        \cline{4-6}
        & & & $\boldsymbol{k\!=\!4}$ & $\boldsymbol{k\!=\!3}$ & $\boldsymbol{k\!=\!2}$ \\
        \hline
        \multirow{4}{*}{Paragraph} & MTurk regular \scriptsize{($n\!=\!5$)} & 0.02 & 0.08 & 0.21 & 0.48 \\
        & MTurk master \scriptsize{($n\!=\!5$)} &  0.18 & 0.35 & 0.53 & 0.73 \\
        \cline{2-6}
        & Prolific \scriptsize{($n\!=\!5$)} & 0.14 & 0.27 & 0.44 & 0.65 \\
        \cline{2-6}
        & Expert \scriptsize{($m\!=\!3$)} &  0.25 & - & - & 0.54 \\
        \cline{2-6}
        \hline
        \multirow{2}{*}{Sentence} & MTurk regular \scriptsize{($n\!=\!3$)} & 0.35 & - & - & 0.71 \\
        & MTurk master \scriptsize{($n\!=\!3$)} & 0.47 & - & - & 0.76 \\
    \end{tabular}
    \label{tab:Jaccard}
\end{table}

\begin{table}
    \centering
    \small
    \caption{Similarity against reference (expert) annotations.}
    \captionshrink
    \begin{tabular}{l|l|c|c|c}
        \textbf{Task variant} & \textbf{Annotator} & $\overline{\boldsymbol{F1}}$ & $\boldsymbol{F1}^{\boldsymbol{\gg}}$ & $\boldsymbol{F1}^{\boldsymbol{\simeq}}$ \\
        \hline
        \multirow{3}{*}{Paragraph-based} & MTurk regular & 0.36 & 0.32 & 0.45 \\
        & MTurk master & \textbf{0.54} & \textbf{0.60} & \textbf{0.61} \\
        \cline{2-5}
        & Prolific & 0.50 & 0.54 & 0.57 \\
        \cline{2-5}
        \hline
        \multirow{2}{*}{Sentence-based} & MTurk regular & 0.31 & 0.33 & 0.34 \\
        & MTurk master & 0.41 & 0.43 & 0.44 \\
    \end{tabular}
    \label{tab:Rouge_variants}
\end{table}

\subsection{Results}

We report on the inter-annotator agreement and similarity against reference annotations on the two topics selected for this preliminary study in Tables~\ref{tab:Jaccard}~and~\ref{tab:Rouge_variants} respectively.

In the paragraph-based variant, we observe better agreement ($J$) between MTurk masters than between Prolific workers, yet there is a big gap between crowd workers and experts. The relative ordering between workers is: MTurk masters $>$ Prolific $>$ MTurk regular, which also holds for the more relaxed version of the agreement measure ($J_k$). We notice that for $J_2$, the agreement between expert annotators is lower than for MTurk masters and Prolific workers; however, there are only 3 experts (vs. 5 crowd workers), hence it is not fair to directly compare these numbers. The generally low agreement scores highlight the difficulty of the task in the paragraph-based form.

On the simplified sentence-based variant, we indeed observe a much higher agreement between MTurk workers.\footnote{Given that MTurk masters outperformed Prolific workers   in the paragraph-based variant, sentence-based annotations are only performed on MTurk.} Also, the differences between regular workers and masters are not as large as in the paragraph-based variant. 
We note that the two task variants (sentence-based and paragraph-based) cannot be compared directly in terms of inter-annotator agreement because the probability of choosing the same snippets by different workers is much higher in a single sentence than in an entire paragraph. 

Table~\ref{tab:Rouge_variants} reports on the quality of worker annotations, with respect to their similarity to the reference (expert) annotations. These results are consistent for all measures and are also in line with the observations made in terms of inter-annotator agreement. Namely, MTurk masters achieve the best results, followed by Prolific workers, and then MTurk regular workers. The same holds for MTurk workers on the sentence-based variant of the task. 
We notice that the absolute scores are much closer for paragraph- and sentence-based annotations than for inter-annotator agreement (with sentence-based performing even slightly better on $F1^{\boldsymbol{\gg}}$ for regular workers). 
Overall, we find that the paragraph-based variant yields higher-quality data than the sentence-based one.

\subsection{Discussion}

Our preliminary exploration of different task designs, platforms, and workers has led us to the conclusion that the highest-quality annotations for this specific task can be collected on the MTurk platform using a paragraph-based task design. 
The main challenge in collecting snippet annotations turned out to be the process of quality control that cannot be automated due to the nature of this task. 
Even for expert annotators, who performed the task attentively, the inter-annotator agreement is low. Therefore, a low similarity between snippets selected by a worker and reference annotations does not imply that the worker did an inferior job. 
Moving forward to collecting annotations at scale, we opt for recruiting a smaller group of crowd workers, using a qualification task, and working closely with them by providing continuous feedback on their work.
\section{Data collection}
\label{sec:data_collection}

This section describes our large-scale data collection effort. For each of the 371 queries in the TREC CAsT 2020 and 2022 datasets, the top 5 passages are annotated by 3 crowd workers, resulting in a total of 1,855 query-passage pairs. 

\subsection{Setup}

The annotation task was released only to a small group of trained crowd workers, who were selected through a qualification task.
The qualification task contained a detailed description of the problem at hand, examples of correct annotations, a quiz, and 10 query-passage pairs to be annotated; 
it was made available to both master and regular MTurk workers to reach a bigger audience.
From the 20 workers that completed the qualification task, we chose 15 that had the highest quality results (independently of their MTurk Master qualification). 
Each worker received feedback on the provided responses and was given an opportunity to ask their own questions about the task. Several rounds of discussion that emerged from the qualification task resulted in an extended set of guidelines addressing the challenging aspects of the annotation task. They contain detailed instructions for crowd workers, a list of tricky cases along with recommendations on how to proceed, a brief description of the problems that we plan to address using the collected data, and a toy examples illustrating how much context should be included in the span. The extended guidelines are made available in the online repository. 

The process of data collection was divided into daily batches and conducted over a period of approximately two weeks. The reason was to both avoid worker fatigue and also to allow for continuous feedback along the way. Each batch contained questions about one specific topic, which amounts to 46 query-passage pairs on average, and was annotated by 3 different workers. 
Workers received \$0.3 for each query-passage pair.
A bonus of \$2 was paid for every batch completed within 24 hours upon release. The total cost of large-scale data collection was \$2.1k.

The training of the annotators did not end at the qualification task but continued throughout the whole data collection process. Crowd workers were provided with feedback after each submitted batch. From each batch, random data samples with low agreement were selected and verified manually by an expert (the main author of the paper). Incorrect data annotations were flagged and discussed individually with crowd workers. After each batch, general comments and suggestions were shared with all workers.
We used Slack as the main communication platform; there, workers could also share challenging cases and benefit collectively from discussions and from expert guidance. The Slack channel was widely used by crowd workers from their own initiative through the whole data annotations process to brainstorm about tricky query-passage pairs and align on their understanding of the task.

\subsection{Statistics} 

\begin{table}[t]
    \small
    \centering
    \caption{Comparison against other datasets.}
    \label{tab:datasets_stats}
    \captionshrink
    \begin{tabular}{l|l|c|c}
       \textbf{Dataset} & \begin{tabular} {@{}l@{}}\textbf{Input}\\\textbf{text} \end{tabular} & \begin{tabular} {@{}l@{}}\textbf{Avg. snippet}\\\textbf{length (tokens)} \end{tabular} & \begin{tabular} {@{}l@{}}\textbf{\#snippets per}\\\textbf{ annotation} \end{tabular} \\
       \hline
        CAsT-snippets & Paragraph & 39.6 & 2.3 \\
        SaaC~\citep{Ren:2021:ToIS} & Top 10 passages & 23.8 & 1.5 \\
        QuaC~\citep{Choi:2018:EMNLP} & Wikipedia article & 14.6 & 1 \\ 
    \end{tabular}

\end{table}

In comparison to the results of the preliminary study (cf. Table~\ref{tab:Jaccard}) on the same set of queries, we find that the inter-annotator agreement ($J$=0.38 and $J_2=$0.62) exceeds even that of expert annotations and the similarity with expert annotations ($\overline{\boldsymbol{F1}}=$0.54) matches those of the best-performing MTurk master workers. These results indicate that the collected data is of high quality and attest to the success of our annotation setup with continuous feedback.

Table~\ref{tab:datasets_stats} provides a comparison against other related datasets. We note that there are not only more snippets annotated for each input text in our dataset, but they are also longer on average, which follows from the information-seeking nature of queries.

We note that there is a number of query-passage pairs where annotators did not find any snippet relevant to the query, despite the passage being labeled as relevant by TREC assessors (77 such passages selected by all three annotators and 111 selected by two of the annotators).

\subsection{Feedback from Crowd Workers}

The close collaboration with crowd workers at every stage of data annotation has revealed several interesting aspects concerning the problem of snippet annotations. One of the most significant challenges was determining the appropriate amount of context to include in each span, striking a balance between conciseness and being self-contained. This issue is closely related to ``conditional responses,'' where the span answers the question only under some specific condition or within a related situation (e.g., a medical condition is mentioned as a symptom, whereas the user was searching for treatment of that condition).  Context also needs to be considered for justification of selected answers, particularly for yes/no responses. Moreover, temporal considerations, such as time mismatches between queries and passages, and the subjectivity of statements in the passage further compounded the challenge.

The second challenge identified by the crowd workers pertains to questions for which only a partial answer can be found in the passage. Deciding whether a span partially answers a question or is only somewhat relevant and should not be selected proved to be highly subjective. Additionally, the crowdsourcing process revealed that even passages with high relevance scores in the ground truth sometimes do not contain the exact answer to the question, resulting in cases of unanswerability.

The third noteworthy observation highlighted by several crowd workers concerns the background knowledge required to select a correct span or determine that the passage does not answer the question. This raises questions about the necessary general/expert knowledge required to annotate responses. Crowd workers worked on batches containing questions from specific areas, and while the TREC CAsT dataset assumes that the information needed to understand the question is included in the conversational context, some contextual knowledge may be missing if the system cannot find a highly relevant passage containing the answer. Moreover, even with access to previous questions and responses within a given topic, crowd workers still encountered challenges when annotating data from topics outside their areas of interest and expertise.

Our task design also included a confidence field for each annotation task, allowing workers to express their level of confidence in the selected spans. We analyzed the collected data to determine whether high confidence levels among workers corresponded to high inter-annotator agreement. Surprisingly, we did not observe any significant relationship between the two measures. We suspect that the confidence scores reported by crowd workers are more closely related to their familiarity with the topic in a given batch.
\section{Conclusions}
\label{sec:conclusions}

We have introduced CAsT-snippets, a high-quality dataset for conversational information seeking containing snippet-level annotations for all queries in the TREC CAsT 2020 and 2022 datasets.
Our annotation effort was informed by a preliminary  study, where we explored various task designs, platforms, and workers pools. Based on the results, we opted for a setup where we closely worked with a pool of highly engaged crowd workers, releasing tasks in daily batches and providing continuous feedback. 

Our direct communication with crowd workers throughout the data annotation process revealed multiple challenges that need to be addressed in conversational response generation: 
(1) Selecting spans for questions when only a partial answer is present is challenging and appears to be highly subjective.
(2) Temporal considerations may exclude some spans as they are not valid answers given the time specified in the query. However, assessing the temporal validity of text may be challenging based solely on short text passages without a larger context.
(3) Passages originating from blogs or comments very often contain subjective opinions. Should such subjective opinions be marked up as answers?
(4) What kind of background knowledge should be assumed when the passage does not contain a direct answer but the answer may be inferred from the text?
(5) How much content is needed for open-ended questions?
(6) When is evidence or additional information needed for a factoid question and when is an entity alone sufficient as an answer?

Our dataset enables the development of answer generation methods that are grounded in relevant snippets in paragraphs as well as allows for the automatic evaluation of the generated answers in terms of completeness; 
a training/test split is provided for such use.

\begin{acks}
This research was supported by the Norwegian Research Center for AI Innovation, NorwAI (Research Council of Norway, project number 309834).
\end{acks}

\balance
\bibliographystyle{ACM-Reference-Format}
\bibliography{cikm2023-snippet_annotation.bib}


\begin{thebibliography}{34}


\ifx \showCODEN    \undefined \def \showCODEN     #1{\unskip}     \fi
\ifx \showDOI      \undefined \def \showDOI       #1{#1}\fi
\ifx \showISBNx    \undefined \def \showISBNx     #1{\unskip}     \fi
\ifx \showISBNxiii \undefined \def \showISBNxiii  #1{\unskip}     \fi
\ifx \showISSN     \undefined \def \showISSN      #1{\unskip}     \fi
\ifx \showLCCN     \undefined \def \showLCCN      #1{\unskip}     \fi
\ifx \shownote     \undefined \def \shownote      #1{#1}          \fi
\ifx \showarticletitle \undefined \def \showarticletitle #1{#1}   \fi
\ifx \showURL      \undefined \def \showURL       {\relax}        \fi
\providecommand\bibfield[2]{#2}
\providecommand\bibinfo[2]{#2}
\providecommand\natexlab[1]{#1}
\providecommand\showeprint[2][]{arXiv:#2}

\bibitem[\protect\citeauthoryear{Baheti, Ritter, and Small}{Baheti et~al\mbox{.}}{2020}]%
        {Baheti:2020:ACL}
\bibfield{author}{\bibinfo{person}{Ashutosh Baheti}, \bibinfo{person}{Alan Ritter}, {and} \bibinfo{person}{Kevin Small}.} \bibinfo{year}{2020}\natexlab{}.
\newblock \showarticletitle{Fluent Response Generation for Conversational Question Answering}. In \bibinfo{booktitle}{\emph{Proceedings of the 58th Annual Meeting of the Association for Computational Linguistics}} \emph{(\bibinfo{series}{ACL '20})}. \bibinfo{pages}{191--207}.
\newblock


\bibitem[\protect\citeauthoryear{Budzianowski, Wen, Tseng, Casanueva, Ultes, Ramadan, and Gasic}{Budzianowski et~al\mbox{.}}{2018}]%
        {Budzianowski:2018:EMNLP}
\bibfield{author}{\bibinfo{person}{Paweł Budzianowski}, \bibinfo{person}{Tsung-Hsien Wen}, \bibinfo{person}{Bo-Hsiang Tseng}, \bibinfo{person}{I{\~n}igo Casanueva}, \bibinfo{person}{Stefan Ultes}, \bibinfo{person}{Osman Ramadan}, {and} \bibinfo{person}{Milica Gasic}.} \bibinfo{year}{2018}\natexlab{}.
\newblock \showarticletitle{{MultiWOZ} - A Large-Scale Multi-Domain {Wizard-of-Oz} Dataset for Task-Oriented Dialogue Modelling}. In \bibinfo{booktitle}{\emph{Findings of the Association for Computational Linguistics: EMNLP 2018}} \emph{(\bibinfo{series}{EMNLP '18})}. \bibinfo{pages}{5016--5026}.
\newblock


\bibitem[\protect\citeauthoryear{Campos, Nguyen, Rosenberg, Song, Gao, Tiwary, Majumder, Deng, and Mitra}{Campos et~al\mbox{.}}{2016}]%
        {Campos:2016:arXiv}
\bibfield{author}{\bibinfo{person}{Daniel~Fernando Campos}, \bibinfo{person}{Tri Nguyen}, \bibinfo{person}{Mir Rosenberg}, \bibinfo{person}{Xia Song}, \bibinfo{person}{Jianfeng Gao}, \bibinfo{person}{Saurabh Tiwary}, \bibinfo{person}{Rangan Majumder}, \bibinfo{person}{Li Deng}, {and} \bibinfo{person}{Bhaskar Mitra}.} \bibinfo{year}{2016}\natexlab{}.
\newblock \bibinfo{title}{{MS MARCO}: A Human Generated {MAchine Reading Comprehension} Dataset}.
\newblock
\newblock
\showeprint[arxiv]{1611.09268}~[cs.CL]


\bibitem[\protect\citeauthoryear{Choi, He, Iyyer, Yatskar, tau Yih, Choi, Liang, and Zettlemoyer}{Choi et~al\mbox{.}}{2018}]%
        {Choi:2018:EMNLP}
\bibfield{author}{\bibinfo{person}{Eunsol Choi}, \bibinfo{person}{He He}, \bibinfo{person}{Mohit Iyyer}, \bibinfo{person}{Mark Yatskar}, \bibinfo{person}{Wen tau Yih}, \bibinfo{person}{Yejin Choi}, \bibinfo{person}{Percy Liang}, {and} \bibinfo{person}{Luke Zettlemoyer}.} \bibinfo{year}{2018}\natexlab{}.
\newblock \showarticletitle{{QuAC}: Question Answering in Context}. In \bibinfo{booktitle}{\emph{Findings of the Association for Computational Linguistics: EMNLP 20}} \emph{(\bibinfo{series}{EMNLP '18})}. \bibinfo{pages}{2174--2184}.
\newblock


\bibitem[\protect\citeauthoryear{Culpepper, Diaz, and Smucker}{Culpepper et~al\mbox{.}}{2018}]%
        {Culpepper:2018:SIGIRForum}
\bibfield{author}{\bibinfo{person}{J~Shane Culpepper}, \bibinfo{person}{Fernando Diaz}, {and} \bibinfo{person}{Mark~D Smucker}.} \bibinfo{year}{2018}\natexlab{}.
\newblock \showarticletitle{Research Frontiers in Information Retrieval: Report from the Third Strategic Workshop on Information Retrieval in {Lorne} ({SWIRL} 2018)}.
\newblock \bibinfo{journal}{\emph{ACM SIGIR Forum}} \bibinfo{volume}{52}, \bibinfo{number}{1} (\bibinfo{year}{2018}), \bibinfo{pages}{34--90}.
\newblock


\bibitem[\protect\citeauthoryear{Dalton, Xiong, and Callan}{Dalton et~al\mbox{.}}{2019}]%
        {Dalton:2019:TREC}
\bibfield{author}{\bibinfo{person}{Jeffrey Dalton}, \bibinfo{person}{Chenyan Xiong}, {and} \bibinfo{person}{Jamie Callan}.} \bibinfo{year}{2019}\natexlab{}.
\newblock \showarticletitle{{TREC} {CAsT} 2019: {The Conversational Assistance} Track Overview}. In \bibinfo{booktitle}{\emph{The Twenty-Eighth Text REtrieval Conference Proceedings}} \emph{(\bibinfo{series}{TREC '19})}.
\newblock


\bibitem[\protect\citeauthoryear{Dalton, Xiong, and Callan}{Dalton et~al\mbox{.}}{2020}]%
        {Dalton:2020:TREC}
\bibfield{author}{\bibinfo{person}{Jeffrey Dalton}, \bibinfo{person}{Chenyan Xiong}, {and} \bibinfo{person}{Jamie Callan}.} \bibinfo{year}{2020}\natexlab{}.
\newblock \showarticletitle{{TREC} {CAsT} 2020: {The Conversational Assistance} Track Overview}. In \bibinfo{booktitle}{\emph{The Twenty-Ninth Text REtrieval Conference Proceedings}} \emph{(\bibinfo{series}{TREC '20})}.
\newblock


\bibitem[\protect\citeauthoryear{Dalton, Xiong, and Callan}{Dalton et~al\mbox{.}}{2021}]%
        {Dalton:2021:TREC}
\bibfield{author}{\bibinfo{person}{Jeffrey Dalton}, \bibinfo{person}{Chenyan Xiong}, {and} \bibinfo{person}{Jamie Callan}.} \bibinfo{year}{2021}\natexlab{}.
\newblock \showarticletitle{{TREC} {CAsT} 2021: {The Conversational Assistance} Track Overview}. In \bibinfo{booktitle}{\emph{The Thirtieth Text REtrieval Conference Proceedings}} \emph{(\bibinfo{series}{TREC '21})}.
\newblock


\bibitem[\protect\citeauthoryear{Daniel, Kucherbaev, Cappiello, Benatallah, and Allahbakhsh}{Daniel et~al\mbox{.}}{2018}]%
        {Daniel:2018:CSUR}
\bibfield{author}{\bibinfo{person}{Florian Daniel}, \bibinfo{person}{Pavel Kucherbaev}, \bibinfo{person}{Cinzia Cappiello}, \bibinfo{person}{Boualem Benatallah}, {and} \bibinfo{person}{Mohammad Allahbakhsh}.} \bibinfo{year}{2018}\natexlab{}.
\newblock \showarticletitle{Quality Control in Crowdsourcing}.
\newblock \bibinfo{journal}{\emph{Comput. Surveys}} \bibinfo{volume}{51}, \bibinfo{number}{1} (\bibinfo{year}{2018}), \bibinfo{pages}{1--40}.
\newblock


\bibitem[\protect\citeauthoryear{Dziri, Kamalloo, Mathewson, and Zaiane}{Dziri et~al\mbox{.}}{2019}]%
        {Dziri:2019:ACL}
\bibfield{author}{\bibinfo{person}{Nouha Dziri}, \bibinfo{person}{Ehsan Kamalloo}, \bibinfo{person}{Kory Mathewson}, {and} \bibinfo{person}{Osmar~R Zaiane}.} \bibinfo{year}{2019}\natexlab{}.
\newblock \showarticletitle{Augmenting Neural Response Generation with Context-Aware Topical Attention}. In \bibinfo{booktitle}{\emph{Proceedings of the First Workshop on NLP for Conversational AI}} \emph{(\bibinfo{series}{ACL '19})}. \bibinfo{pages}{18--31}.
\newblock


\bibitem[\protect\citeauthoryear{Eickhoff}{Eickhoff}{2018}]%
        {Eickhoff:2018:WSDM}
\bibfield{author}{\bibinfo{person}{Carsten Eickhoff}.} \bibinfo{year}{2018}\natexlab{}.
\newblock \showarticletitle{Cognitive Biases in Crowdsourcing}. In \bibinfo{booktitle}{\emph{Proceedings of the 11th ACM International Conference on Web Search and Data Mining}} \emph{(\bibinfo{series}{WSDM '18})}. \bibinfo{pages}{162--170}.
\newblock


\bibitem[\protect\citeauthoryear{Fabbri, Kryscinski, McCann, Socher, and Radev}{Fabbri et~al\mbox{.}}{2020}]%
        {Fabbri:2020:ACL}
\bibfield{author}{\bibinfo{person}{A.~R. Fabbri}, \bibinfo{person}{Wojciech Kryscinski}, \bibinfo{person}{Bryan McCann}, \bibinfo{person}{Richard Socher}, {and} \bibinfo{person}{Dragomir~R. Radev}.} \bibinfo{year}{2020}\natexlab{}.
\newblock \showarticletitle{{SummEval}: Re-evaluating Summarization Evaluation}.
\newblock \bibinfo{journal}{\emph{Transactions of the Association for Computational Linguistics}}  \bibinfo{volume}{9} (\bibinfo{year}{2020}), \bibinfo{pages}{391--409}.
\newblock


\bibitem[\protect\citeauthoryear{Gadiraju, Demartini, Kawase, and Dietze}{Gadiraju et~al\mbox{.}}{2015}]%
        {Gadiraju:2015:IEEE}
\bibfield{author}{\bibinfo{person}{Ujwal Gadiraju}, \bibinfo{person}{Gianluca Demartini}, \bibinfo{person}{Ricardo Kawase}, {and} \bibinfo{person}{Stefan Dietze}.} \bibinfo{year}{2015}\natexlab{}.
\newblock \showarticletitle{Human Beyond the Machine: Challenges and Opportunities of Microtask Crowdsourcing}.
\newblock \bibinfo{journal}{\emph{IEEE Intelligent Systems}}  \bibinfo{volume}{30} (\bibinfo{year}{2015}), \bibinfo{pages}{81--85}.
\newblock


\bibitem[\protect\citeauthoryear{Iskender, Schaefer, Polzehl, and M{\"o}ller}{Iskender et~al\mbox{.}}{2021}]%
        {Iskender:2021:NLDB}
\bibfield{author}{\bibinfo{person}{Neslihan Iskender}, \bibinfo{person}{Robin Schaefer}, \bibinfo{person}{Tim Polzehl}, {and} \bibinfo{person}{Sebastian M{\"o}ller}.} \bibinfo{year}{2021}\natexlab{}.
\newblock \showarticletitle{Argument Mining in Tweets: Comparing Crowd and Expert Annotations for Automated Claim and Evidence Detection}. In \bibinfo{booktitle}{\emph{International Conference on Applications of Natural Language to Data Bases}} \emph{(\bibinfo{series}{NLDB '21})}. \bibinfo{pages}{275--288}.
\newblock


\bibitem[\protect\citeauthoryear{Ji, Lee, Frieske, Yu, Su, Xu, Ishii, Bang, Madotto, and Fung}{Ji et~al\mbox{.}}{2022}]%
        {Ji:2022:arxiv}
\bibfield{author}{\bibinfo{person}{Ziwei Ji}, \bibinfo{person}{Nayeon Lee}, \bibinfo{person}{Rita Frieske}, \bibinfo{person}{Tiezheng Yu}, \bibinfo{person}{Dan Su}, \bibinfo{person}{Yan Xu}, \bibinfo{person}{Etsuko Ishii}, \bibinfo{person}{Yejin Bang}, \bibinfo{person}{Andrea Madotto}, {and} \bibinfo{person}{Pascale Fung}.} \bibinfo{year}{2022}\natexlab{}.
\newblock \bibinfo{title}{Survey of Hallucination in Natural Language Generation}.
\newblock
\newblock
\showeprint[arxiv]{2202.03629}~[cs.CL]


\bibitem[\protect\citeauthoryear{Kocisk{\'y}, Schwarz, Blunsom, Dyer, Hermann, Melis, and Grefenstette}{Kocisk{\'y} et~al\mbox{.}}{2017}]%
        {Kocisk:2017:TACL}
\bibfield{author}{\bibinfo{person}{Tom{\'a}s Kocisk{\'y}}, \bibinfo{person}{Jonathan Schwarz}, \bibinfo{person}{Phil Blunsom}, \bibinfo{person}{Chris Dyer}, \bibinfo{person}{Karl~Moritz Hermann}, \bibinfo{person}{G{\'a}bor Melis}, {and} \bibinfo{person}{Edward Grefenstette}.} \bibinfo{year}{2017}\natexlab{}.
\newblock \showarticletitle{The {NarrativeQA} Reading Comprehension Challenge}.
\newblock \bibinfo{journal}{\emph{Transactions of the Association for Computational Linguistics}}  \bibinfo{volume}{6} (\bibinfo{year}{2017}), \bibinfo{pages}{317--328}.
\newblock


\bibitem[\protect\citeauthoryear{Lin}{Lin}{2004}]%
        {Lin:2004:ACL}
\bibfield{author}{\bibinfo{person}{Chin-Yew Lin}.} \bibinfo{year}{2004}\natexlab{}.
\newblock \showarticletitle{{ROUGE}: A Package for Automatic Evaluation of Summaries}. In \bibinfo{booktitle}{\emph{Proceedings of Workshop on Text Summarization Branches Out on Annual Meeting of the Association for Computational Linguistics}} \emph{(\bibinfo{series}{ACL '04})}. \bibinfo{pages}{74--81}.
\newblock


\bibitem[\protect\citeauthoryear{Lin, Yang, Nogueira, Tsai, Wang, and Lin}{Lin et~al\mbox{.}}{2021}]%
        {Lin:2021:TOIS}
\bibfield{author}{\bibinfo{person}{Sheng-Chieh Lin}, \bibinfo{person}{Jheng-Hong Yang}, \bibinfo{person}{Rodrigo Nogueira}, \bibinfo{person}{Ming-Feng Tsai}, \bibinfo{person}{Chuan-Ju Wang}, {and} \bibinfo{person}{Jimmy Lin}.} \bibinfo{year}{2021}\natexlab{}.
\newblock \showarticletitle{Multi-stage Conversational Passage Retrieval: An Approach to Fusing Term Importance Estimation and Neural Query Rewriting}.
\newblock \bibinfo{journal}{\emph{ACM Transactions on Information Systems}} \bibinfo{volume}{39}, \bibinfo{number}{4} (\bibinfo{year}{2021}), \bibinfo{pages}{1--29}.
\newblock


\bibitem[\protect\citeauthoryear{Lippe, Ren, Haned, Voorn, and de~Rijke}{Lippe et~al\mbox{.}}{2020}]%
        {Lippe:2020:arXiv}
\bibfield{author}{\bibinfo{person}{Phillip Lippe}, \bibinfo{person}{Pengjie Ren}, \bibinfo{person}{Hinda Haned}, \bibinfo{person}{Bart Voorn}, {and} \bibinfo{person}{M. de Rijke}.} \bibinfo{year}{2020}\natexlab{}.
\newblock \bibinfo{title}{Diversifying Task-oriented Dialogue Response Generation with Prototype Guided Paraphrasing}.
\newblock
\newblock
\showeprint[arxiv]{2008.03391}~[cs.CL]


\bibitem[\protect\citeauthoryear{Luan, Eisenstein, Toutanova, and Collins}{Luan et~al\mbox{.}}{2021}]%
        {Luan:2021:MIT}
\bibfield{author}{\bibinfo{person}{Yi Luan}, \bibinfo{person}{Jacob Eisenstein}, \bibinfo{person}{Kristina Toutanova}, {and} \bibinfo{person}{Michael Collins}.} \bibinfo{year}{2021}\natexlab{}.
\newblock \showarticletitle{Sparse, Dense, and Attentional Representations for Text Retrieval}.
\newblock \bibinfo{journal}{\emph{Transactions of the Association for Computational Linguistics}}  \bibinfo{volume}{9} (\bibinfo{year}{2021}), \bibinfo{pages}{329--345}.
\newblock


\bibitem[\protect\citeauthoryear{Owoicho, Dalton, Aliannejad, Azzopardi, Trippas, and Vakulenko}{Owoicho et~al\mbox{.}}{2022}]%
        {Owoicho:2022:TREC}
\bibfield{author}{\bibinfo{person}{Paul Owoicho}, \bibinfo{person}{Jeffrey Dalton}, \bibinfo{person}{Mohammad Aliannejad}, \bibinfo{person}{Leif Azzopardi}, \bibinfo{person}{Johanne~R. Trippas}, {and} \bibinfo{person}{Svitlana Vakulenko}.} \bibinfo{year}{2022}\natexlab{}.
\newblock \showarticletitle{{TREC} {CAsT} 2022: Going Beyond User Ask and System Retrieve with Initiative and Response Generation}. In \bibinfo{booktitle}{\emph{The Thirty-First Text REtrieval Conference Proceedings}} \emph{(\bibinfo{series}{TREC '22})}.
\newblock


\bibitem[\protect\citeauthoryear{Papineni, Roukos, Ward, and Zhu}{Papineni et~al\mbox{.}}{2002}]%
        {Papineni:2002:ACL}
\bibfield{author}{\bibinfo{person}{Kishore Papineni}, \bibinfo{person}{Salim Roukos}, \bibinfo{person}{Todd Ward}, {and} \bibinfo{person}{Wei-Jing Zhu}.} \bibinfo{year}{2002}\natexlab{}.
\newblock \showarticletitle{{BLEU}: a Method for Automatic Evaluation of Machine Translation}. In \bibinfo{booktitle}{\emph{Proceedings of the 40th Annual Meeting of the Association for Computational Linguistics}} \emph{(\bibinfo{series}{ACL '02})}. \bibinfo{pages}{311--318}.
\newblock


\bibitem[\protect\citeauthoryear{Pavlu, Rajput, Golbus, and Aslam}{Pavlu et~al\mbox{.}}{2012}]%
        {Pavlu:2012:WSDM}
\bibfield{author}{\bibinfo{person}{Virgil Pavlu}, \bibinfo{person}{Shahzad Rajput}, \bibinfo{person}{Peter~B. Golbus}, {and} \bibinfo{person}{Javed~A. Aslam}.} \bibinfo{year}{2012}\natexlab{}.
\newblock \showarticletitle{{IR} System Evaluation Using Nugget-based Test Collections}. In \bibinfo{booktitle}{\emph{Proceedings of the 5th ACM International Conference on Web Search and Data Mining}} \emph{(\bibinfo{series}{WSDM '12})}. \bibinfo{pages}{393--402}.
\newblock


\bibitem[\protect\citeauthoryear{Pei, Ren, Monz, and de~Rijke}{Pei et~al\mbox{.}}{2019}]%
        {Pei:2019:ECAI}
\bibfield{author}{\bibinfo{person}{Jiahuan Pei}, \bibinfo{person}{Pengjie Ren}, \bibinfo{person}{Christof Monz}, {and} \bibinfo{person}{M. de Rijke}.} \bibinfo{year}{2019}\natexlab{}.
\newblock \showarticletitle{Retrospective and Prospective Mixture-of-Generators for Task-oriented Dialogue Response Generation}. In \bibinfo{booktitle}{\emph{European Conference on Artificial Intelligence}} \emph{(\bibinfo{series}{ECAI '19})}.
\newblock


\bibitem[\protect\citeauthoryear{Rajpurkar, Zhang, Lopyrev, and Liang}{Rajpurkar et~al\mbox{.}}{2016}]%
        {Rajpurkar:2016:EMNLP}
\bibfield{author}{\bibinfo{person}{Pranav Rajpurkar}, \bibinfo{person}{Jian Zhang}, \bibinfo{person}{Konstantin Lopyrev}, {and} \bibinfo{person}{Percy Liang}.} \bibinfo{year}{2016}\natexlab{}.
\newblock \showarticletitle{{SQuAD}: 100,000+ Questions for Machine Comprehension of Text}. In \bibinfo{booktitle}{\emph{Findings of the Association for Computational Linguistics: EMNLP 2016}} \emph{(\bibinfo{series}{EMNLP '16})}. \bibinfo{pages}{2383--2392}.
\newblock


\bibitem[\protect\citeauthoryear{Ren, Chen, Ren, Kanoulas, Monz, and de~Rijke}{Ren et~al\mbox{.}}{2021}]%
        {Ren:2021:ToIS}
\bibfield{author}{\bibinfo{person}{Pengjie Ren}, \bibinfo{person}{Zhumin Chen}, \bibinfo{person}{Zhaochun Ren}, \bibinfo{person}{E. Kanoulas}, \bibinfo{person}{Christof Monz}, {and} \bibinfo{person}{M. de Rijke}.} \bibinfo{year}{2021}\natexlab{}.
\newblock \showarticletitle{Conversations with Search Engines: {SERP-based} Conversational Response Generation}.
\newblock \bibinfo{journal}{\emph{ACM Transactions on Information Systems}} \bibinfo{volume}{39}, \bibinfo{number}{4} (\bibinfo{year}{2021}), \bibinfo{pages}{1--29}.
\newblock


\bibitem[\protect\citeauthoryear{Tan, Wei, Yang, Lv, and Zhou}{Tan et~al\mbox{.}}{2017}]%
        {Tan:2017:arXiv}
\bibfield{author}{\bibinfo{person}{Chuanqi Tan}, \bibinfo{person}{Furu Wei}, \bibinfo{person}{Nan Yang}, \bibinfo{person}{Weifeng Lv}, {and} \bibinfo{person}{M. Zhou}.} \bibinfo{year}{2017}\natexlab{}.
\newblock \bibinfo{title}{{S-Net}: From Answer Extraction to Answer Generation for Machine Reading Comprehension}.
\newblock
\newblock
\showeprint[arxiv]{1706.04815}~[cs.CL]


\bibitem[\protect\citeauthoryear{Trippas, Spina, Cavedon, Joho, and Sanderson}{Trippas et~al\mbox{.}}{2018}]%
        {Trippas:2018:CHIIR}
\bibfield{author}{\bibinfo{person}{Johanne~R. Trippas}, \bibinfo{person}{Damiano Spina}, \bibinfo{person}{Lawrence Cavedon}, \bibinfo{person}{Hideo Joho}, {and} \bibinfo{person}{Mark Sanderson}.} \bibinfo{year}{2018}\natexlab{}.
\newblock \showarticletitle{Informing the Design of Spoken Conversational Search: Perspective Paper}. In \bibinfo{booktitle}{\emph{Proceedings of the 2018 Conference on Human Information Interaction \& Retrieval}} \emph{(\bibinfo{series}{CHIIR '18})}. \bibinfo{pages}{32--41}.
\newblock


\bibitem[\protect\citeauthoryear{Trippas, Spina, Thomas, Sanderson, Joho, and Cavedon}{Trippas et~al\mbox{.}}{2019}]%
        {Trippas:2019:IPM}
\bibfield{author}{\bibinfo{person}{Johanne~R. Trippas}, \bibinfo{person}{Damiano Spina}, \bibinfo{person}{Paul Thomas}, \bibinfo{person}{Mark Sanderson}, \bibinfo{person}{Hideo Joho}, {and} \bibinfo{person}{Lawrence Cavedon}.} \bibinfo{year}{2019}\natexlab{}.
\newblock \showarticletitle{Towards a Model for Spoken Conversational Search}.
\newblock \bibinfo{journal}{\emph{Information Processing \& Management}} \bibinfo{volume}{57}, \bibinfo{number}{2} (\bibinfo{year}{2019}), \bibinfo{pages}{102--162}.
\newblock


\bibitem[\protect\citeauthoryear{Vakharia and Lease}{Vakharia and Lease}{2013}]%
        {Vakharia:2013:arXiv}
\bibfield{author}{\bibinfo{person}{Donna Vakharia} {and} \bibinfo{person}{Matthew Lease}.} \bibinfo{year}{2013}\natexlab{}.
\newblock \bibinfo{title}{Beyond {AMT}: An Analysis of Crowd Work Platforms}.
\newblock
\newblock
\showeprint[arxiv]{1310.1672}~[cs.CY]


\bibitem[\protect\citeauthoryear{Vakulenko, Longpre, Tu, and Anantha}{Vakulenko et~al\mbox{.}}{2021a}]%
        {Vakulenko:2021:WSDM}
\bibfield{author}{\bibinfo{person}{Svitlana Vakulenko}, \bibinfo{person}{Shayne Longpre}, \bibinfo{person}{Zhucheng Tu}, {and} \bibinfo{person}{Raviteja Anantha}.} \bibinfo{year}{2021}\natexlab{a}.
\newblock \showarticletitle{Question Rewriting for Conversational Question Answering}. In \bibinfo{booktitle}{\emph{Proceedings of the 14th ACM International Conference on Web Search and Data Mining}} \emph{(\bibinfo{series}{WSDM '21})}. \bibinfo{pages}{355--363}.
\newblock


\bibitem[\protect\citeauthoryear{Vakulenko, Voskarides, Tu, and Longpre}{Vakulenko et~al\mbox{.}}{2021b}]%
        {Vakulenko:2021:ECIR}
\bibfield{author}{\bibinfo{person}{Svitlana Vakulenko}, \bibinfo{person}{Nikos Voskarides}, \bibinfo{person}{Zhucheng Tu}, {and} \bibinfo{person}{Shayne Longpre}.} \bibinfo{year}{2021}\natexlab{b}.
\newblock \showarticletitle{A Comparison of Question Rewriting Methods for Conversational Passage Retrieval}. In \bibinfo{booktitle}{\emph{European Conference on Information Retrieval}} \emph{(\bibinfo{series}{ECIR '21})}. \bibinfo{pages}{418--424}.
\newblock


\bibitem[\protect\citeauthoryear{Xing, Wu, Wu, Liu, Huang, Zhou, and Ma}{Xing et~al\mbox{.}}{2016}]%
        {Xing:2016:AAAI}
\bibfield{author}{\bibinfo{person}{Chen Xing}, \bibinfo{person}{Wei Wu}, \bibinfo{person}{Yu Wu}, \bibinfo{person}{Jie Liu}, \bibinfo{person}{Yalou Huang}, \bibinfo{person}{M. Zhou}, {and} \bibinfo{person}{Wei-Ying Ma}.} \bibinfo{year}{2016}\natexlab{}.
\newblock \showarticletitle{Topic Aware Neural Response Generation}. In \bibinfo{booktitle}{\emph{Proceedings of the AAAI Conference on Artificial Intelligence}} \emph{(\bibinfo{series}{AAAI '16})}. \bibinfo{pages}{3351--3357}.
\newblock


\bibitem[\protect\citeauthoryear{Zhou, Yang, Wei, Huang, Zhou, and Zhao}{Zhou et~al\mbox{.}}{2018}]%
        {Zhou:2018:ACL}
\bibfield{author}{\bibinfo{person}{Qingyu Zhou}, \bibinfo{person}{Nan Yang}, \bibinfo{person}{Furu Wei}, \bibinfo{person}{Shaohan Huang}, \bibinfo{person}{M. Zhou}, {and} \bibinfo{person}{Tiejun Zhao}.} \bibinfo{year}{2018}\natexlab{}.
\newblock \showarticletitle{Neural Document Summarization by Jointly Learning to Score and Select Sentences}. In \bibinfo{booktitle}{\emph{Proceedings of the 56th Annual Meeting of the Association for Computational Linguistics}} \emph{(\bibinfo{series}{ACL '18})}. \bibinfo{pages}{654--663}.
\newblock


\end{thebibliography}

\end{document}